\documentclass[aps,pre,
 twocolumn,
 notitlepage,
 showpacs,floatfix,nofootinbib,
 superscriptaddress
 ]{revtex4-1}

 \usepackage{subfigure}
 \usepackage{amssymb}
 \usepackage{amsfonts}
 \usepackage{amsmath}
 \usepackage{amsthm}
 \usepackage{epsfig}
 \usepackage{float}
 \usepackage[usenames,dvipsnames]{color}
 \usepackage[latin1]{inputenc}

 \usepackage{graphics,graphicx,xcolor,subfigure}

 \usepackage[hidelinks,unicode=true]{hyperref}
 \hypersetup{colorlinks=true,
 	linkcolor=blue,
 	urlcolor=blue,
 	citecolor=blue,
 	pdfhighlight=/N
 }
 
 \usepackage{comment}
 \usepackage[normalem]{ulem}
 \usepackage{microtype}

\newcommand{\av}[1]{\langle {#1} \rangle}

\newcommand{\kmax}{k_\text{max}}
\newcommand{\kmin}{k_\text{min}}
\newcommand{\kc}{k_\text{c}}

\newcommand{\fperc}{f_\text{c}^\text{per}}
\newcommand{\fimmu}{f_\text{c}^\text{imm}}
\newcommand{\fimmc}{f_\text{c}^\text{imm}}

\begin{document}

\title{Non massive immunization to contain spreading on complex networks}

\author{Guilherme S. Costa}
\affiliation{Departamento de F\'{\i}sica, Universidade Federal de Vi\c{c}osa, 36570-900 Vi\c{c}osa, Minas Gerais, Brazil}

\author{Silvio C. Ferreira}
\affiliation{Departamento de F\'{\i}sica, Universidade Federal de Vi\c{c}osa, 36570-900 Vi\c{c}osa, Minas Gerais, Brazil}
\affiliation{National Institute of Science and Technology for Complex Systems, 22290-180, Rio de Janeiro, Brazil}

\begin{abstract} 
Optimal strategies for epidemic containment are focused on dismantling the
contact network  through effective immunization  with minimal costs. However,
network fragmentation is seldom accessible in practice and may present extreme
side effects. In this work, we investigate  the epidemic containment immunizing
population fractions far below the percolation threshold. We report that
moderate and weakly supervised immunizations can lead to finite epidemic thresholds
of the susceptible-infected-susceptible  model on scale-free networks by
changing the nature of the transition from a specific-motif to a collectively
driven process. Both pruning of efficient spreaders and increasing of their
mutual separation are necessary for a collective activation. Fractions of
immunized vertices needed to eradicate the epidemics which are  much smaller than the
percolation thresholds were observed for a broad spectrum of real networks
considering targeted or acquaintance immunization strategies. Our work
contributes for the construction of optimal containment, preserving network
functionality through non massive and viable immunization strategies.
\end{abstract}


\maketitle

\section{Introduction}

Modern societies are strongly regulated by networked systems such as
face-to-face~\cite{Cattuto2010,Stehle2011,Liljeros2001} and remote
social~\cite{Ebel2002} interactions, transportation
infrastructures~\cite{Colizza2006,Balcan2009}, communication
networks~\cite{Broder2000, Vazquez2002, Nanavati2006}, and so on. These
substrates can also be underlying structures through which threats propagate,
such as the spreading of contagious diseases~\cite{anderson1992infectious,
	Pastor-Satorras2015}, computer viruses~\cite{Pastor-Satorras2001}, and fake
news~\cite{Shu2017,Lazer2018,Tornberg2018}. Therefore understanding
immunization or knockout (depending on the context) strategies is
fundamental~\cite{Wang2016} either by the necessity to prevent a menace
propagation, as in a contagion disease or criminal
network~\cite{DaCunha2018,Latora2004}, or to protect vital components, as in
communication~\cite{Broder2000,Nanavati2006} and
transportation~\cite{Colizza2006,Balcan2009} infrastructures. Epidemic models
can be interpreted as generic spreading processes~\cite{Pastor-Satorras2015} and
we hereafter adopt epidemiology jargons without loss of generality.

Containment methods are frequently associated with the percolation
analysis~\cite{Dorogovtsev2008}, in which the immunization of nodes or edges
would lead to the  fragmentation of the transmission network into small components,
hindering the spreading~\cite{Cohen2003,Chen2008,Gallos2007}. One remarkable
property of scale-free (SF) networks, represented by degree distributions with
power-law tails in the form $P(k)\sim k^{-\gamma}$ with degree exponent
$2<\gamma<3$, is their resilience to random immunization that is ineffective to
fragment the network~\cite{Cohen2001,Albert2002}. However, targeted immunization
of a fraction $f<1$ of the  most central nodes can dismantle SF
networks~\cite{Wang2016,Albert2002}. Immunization based on degree and
betweenness~\cite{BOCCALETTI2006} using global properties of the network  are
commonly used~\cite{Cohen2001,Pastor-Satorras2002,Holme2002};
see~\cite{Wang2016} for a survey of strategies. {A drawback of global methods is
	that knowledge about properties of all nodes is frequently not accessible due
	 to either computational limitations or lack of information of the network
	topology}. An efficient and flexible alternative considers acquaintances of
randomly selected nodes based only on local information~\cite{Cohen2003}. This
approach, in which a neighbor of a randomly selected vertex is chosen to be
immunized, is grounded on the fact that acquaintances are, on average, more
central than randomly selected nodes~\cite{Newman2002}.

Network fragmentation into subextensive components will certainly prevent
large-scale epidemic spreading~\cite{Cohen2003,Chen2008}. However, if the
percolation threshold is high, network fragmentation can be an impracticable
attitude due to the costs and harmful side effects. So, how do epidemic
processes evolve on moderately  immunized (no fragmented) networks? And which
fraction of immunization is needed to prevent the epidemic spreading in
comparison with the percolation threshold? Despite of explicit analyses of
epidemic spreading on immunized networks~\cite{Pastor-Satorras2002, Dezso2002,
	Holme2004, Gomez-Gardenes2006, Matamalas2018}, these issues have not been
addressed thoroughly to the best of our knowledge.

Consider the SIS model~\cite{anderson1992infectious}  on the top of complex
networks~\cite{Pastor-Satorras2001, Pastor-Satorras2015} where nodes can be in
one of two states: \textit{infected}, which become spontaneously susceptible
with rate $\mu=1$, or \textit{susceptible}, which can be infected by each  of
their infected contacts with rate $\lambda$. The epidemic threshold
$\lambda_\text{c}$ determines the infection rate above which the epidemics can
thrive indefinitely in an extensive portion of the network. A remarkable feature
of the SIS model is its null epidemic threshold for SF networks as the size
$N\rightarrow\infty$~\cite{Pastor-Satorras2001, Moreno2002, Chatterjee2009}
meaning that the epidemics always reaches a finite fraction of the network
irrespective of the value of $\lambda$. The activation mechanisms of epidemic
process and, in particular, of SIS can be quite tricky to
analyze~\cite{Castellano2018,Kitsak2010,Sander2016,Castellano2012}. We can
classify the activation into motif-driven and collective
processes~\cite{Sander2016,Cota2018a}. In the former, a subextensive fraction is
responsible for the triggering the epidemics and spreading it out to the rest of
the network infecting an extensive fraction the population and the epidemic
threshold vanishes as $N\rightarrow\infty$. This is the case of the SIS model on
power-law networks for which activation can be triggered by either hubs or  a
densely connected subgraph given by the maximal index of a k-core decomposition,
depending on the degree exponent~\cite{Castellano2012}. In the case of
collective activation, an extensive part of the network is directly
involved~\cite{Sander2016,Cota2018a}. This happens, for example, in the Harris
contact process for any value of the degree exponent~\cite{Cota2018a} and in the
susceptible-infected-recovered-susceptible (SIRS) model for
$\gamma>3$~\cite{Sander2016}. In the SIRS model, the infected individual stays
for a while in an immunized state before becoming susceptible again (wanning
immunity)~\cite{anderson1992infectious}.

Since the SIS model possesses a fluctuating active steady state, its connection with
percolation is  not immediate as in the susceptible-infected-recovered (SIR)
model~\cite{Cohen2003,Schneider2011}, in which an immunized node becomes
recovered and cannot be reinfected. Whereas random immunization is ineffective,
targeted strategies~\cite{Pastor-Satorras2002, Dezso2002} can lead to a finite
epidemic threshold in SF networks through the immunization of the most connected
nodes in both SIS and SIR models. Acquaintance immunization can also lead to an
finite epidemic threshold of the  SIR model on SF networks~\cite{Cohen2003}. Recently,
Matamalas \textit{et al}. ~\cite{Matamalas2018} considered removal of edges with
the highest probability to  transmit the disease considering a discrete-time
version of the pairwise mean-field theory for the SIS model~\cite{Mata2013}.
This method successfully promoted epidemic containment preserving a connected
giant component. However, this is global approach prone to aforementioned
difficulties of applications in large networks.

In this paper, we push forward  this field investigating the evolution of the
susceptible-infected-susceptible (SIS) model on synthetic and real networks
where a fraction of the nodes far below the percolation threshold is immunized.
We consider distinct immunization strategies, including global and local
methods. We report that a non massive and weakly supervised immunization can
promote containment by altering the nature of the epidemic transition from a
specific-motif to a collectively driven activation, permitting that other
processes remain functioning after network immunization. We also show that
immunized networks are structurally different from their randomized counterparts
and a finite epidemic threshold can emerge even when its randomized version
still has a vanishing one.

The rest of this paper is organized as follows: In Section~\ref{sec:immu}, we
discuss the immunization strategies and the structure of synthetic networks
resulting from them. We present the investigation of the epidemic threshold of
the SIS model on immunized synthetic networks in Sec.~\ref{sec:sinSIS}. Effects
of immunization on a collection real networks are presented in
Sec.~\ref{sec:threReal}. In Section \ref{sec:conclu}, we summarize the findings
of this paper and draw our concluding remarks. Appendix~\ref{app:strucreal}
presents a summary of the real networks used in the current work while
Appendices~\ref{app:numer} and \ref{app:mft} complement the paper with some
computational and analytical methods  used throughout the paper.

\section{Immunized network analyses}
\label{sec:immu}
Consider an initially connected network with $N$ nodes, in which a fraction $f$
will be immunized, which means that the vertex and all edges connected to it
will be removed. We considered only adaptive
methods~\cite{Holme2002,Schneider2011}, in which the network properties are
recalculated every time a node is immunized. In the targeted immunization (TgI),
each step corresponds to immunize the most connected vertex of the network. In
acquaintance immunization (AcI),  a vertex and one of its nearest-neighbors
(acquaintances) are sequentially selected at random, being  a local
strategy~\cite{Cohen2003,Holme2004}. The neighbor is immunized with probability
\begin{equation}
\Phi(k)=\frac{\av{k}^s}{\av{k}^s+k^s},
\label{eq:Phik}
\end{equation}
where the degree $k$ is the number of nonimmunized nearest-neighbors  of the
vertex to be immunized, $\av{k}$ the average degree of the original network, and
$s$ is a parameter. If $s=0$, it is the adaptive version of the acquaintance
strategy of Ref.~\cite{Cohen2003}. If $s>0$, hubs are protected having a smaller
probability to be immunized while $s<0$ implies that hubs are selected
preferentially resembling TgI. Here, we present results for $s=0$ and $s=+1/2$,
hereafter, called AcI and AcI with hub protection (AcI-HP), respectively. The
latter can be considered a weakly supervised strategy due to its limited
capacity to determine the most efficient spreaders.
\begin{figure}[hbt]
	\centering
	\includegraphics[width=0.95\columnwidth]{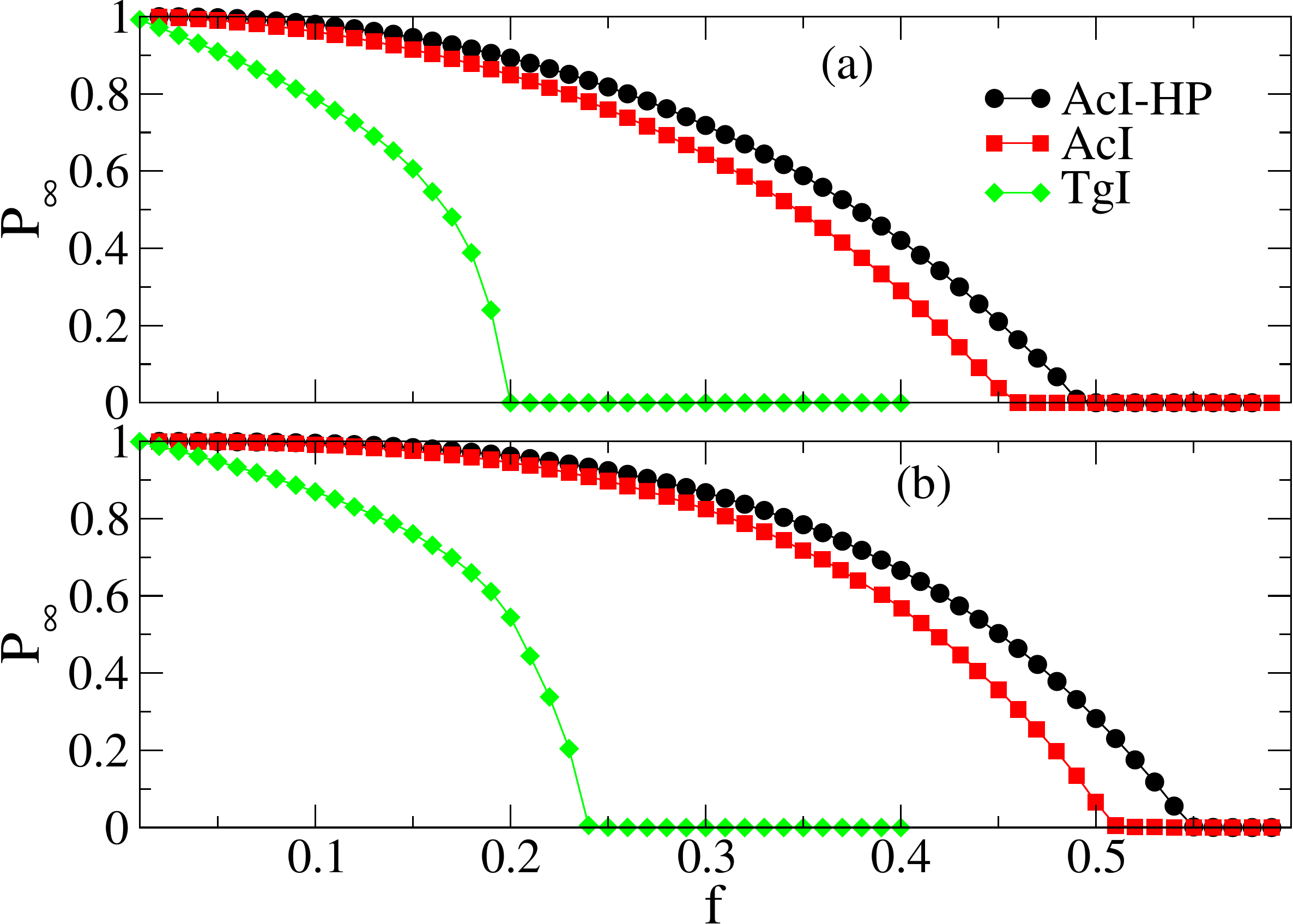}
	\caption{Percolation analysis on synthetic SF networks with $N=10^7$
		nodes considering three immunization methods defined in the main paper. Two
		values of the degree exponent  (a) $\gamma = 2.3$ and (b) $2.8$ are presented.
		The curves correspond to averages over 100 networks with one realization of
		immunization per network. Abscissas are the same in both plots.}
	\label{fig:perc}
\end{figure}

We consider  synthetic networks generated with the uncorrelated configuration
model (UCM)~\cite{Catanzaro2005} using a minimal degree $\kmin=3$ and an upper
cutoff $\kc=\sqrt{N}$. We performed percolation analyses to determine whether
the fraction of immunized vertices $f$ fragments or not the network into small
components. Figure~\ref{fig:perc} shows the fraction of nonimmunized nodes
$P_\infty$ which belong to the largest connected component
(LCC)~\cite{Newman2014} as a function of $f$ for synthetic SF networks of size
$N = 10^7$ and two values of the degree exponent $\gamma<3$. The percolation
thresholds $\fperc$, separating phases with an extensive ($P_\infty>0$) and
subextensive ($P_\infty=0$) LCC, are given in Table~\ref{tab:pth}. To deal with
finite networks we assume that a relative size of the LCC below $10^{-3}$
corresponds to the percolation thresholds, being the results little sensitive to
this choice. Figure 1 shows that the LCC  for $f \approx \fperc/5$ corresponds
to more than  90\% of the nonimmunized vertices in all cases, being nearly 100\%
for AcI and AcI-HP methods.

 \begin{figure}[tbh]
 	\centering
 	\includegraphics[height=0.85\linewidth,angle=90]{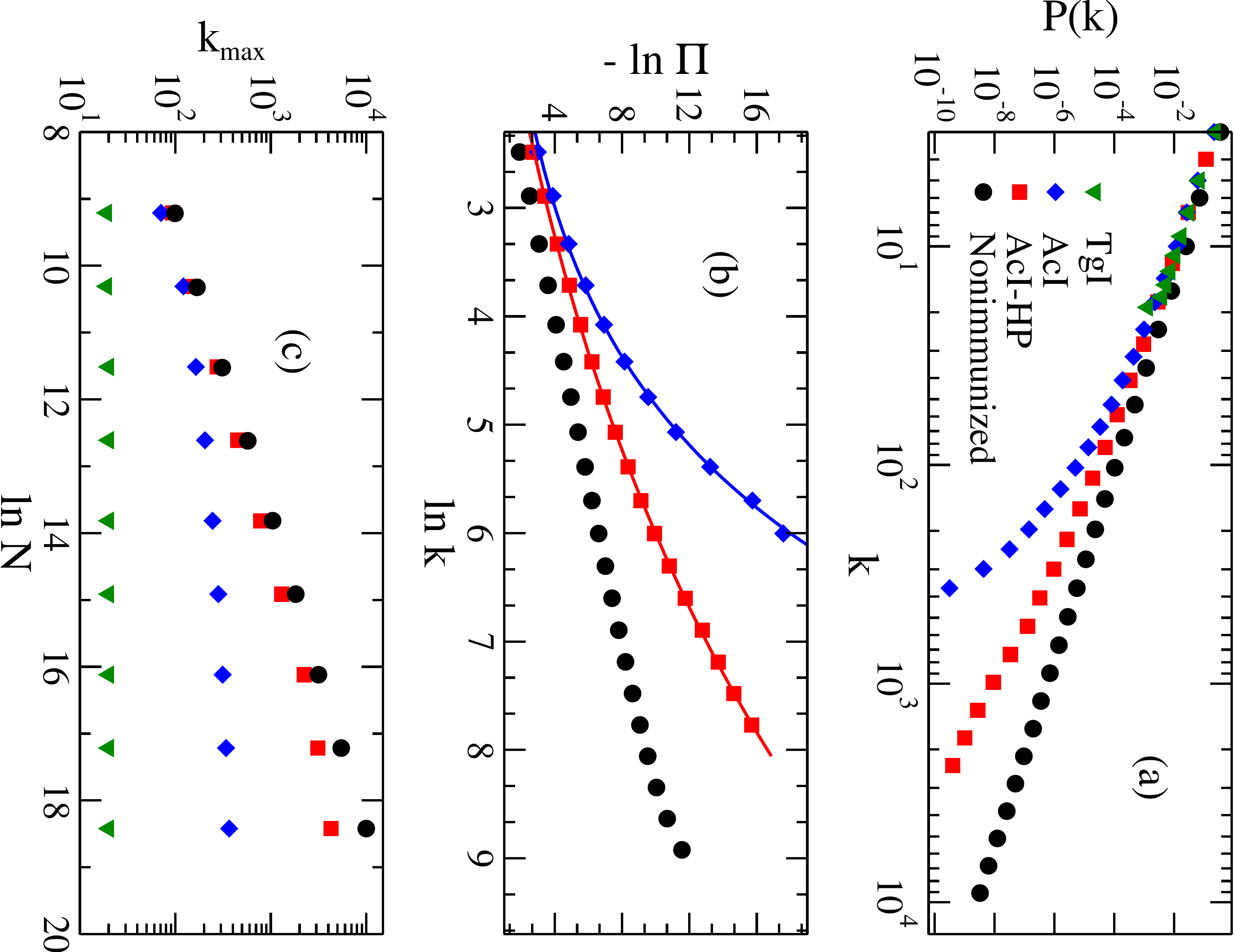}	
 	\caption{Structural analysis of UCM networks with degree exponent $\gamma=2.3$
 		immunized  using different methods with $f=\fperc/5$. (a) Degree and (b) tail
 		distributions  for $N=10^8$. (c) Average largest degree as function of the
 		logarithm of the network size. Distribution curves were smoothed by a
 		logarithm binning~\cite{barabasi2016network}.  In panel (b), solid lines are
 		nonlinear regressions using stretched exponential.}
 	\label{fig:stru_summ23}
 \end{figure}

Basic structural properties of immunized networks considering $f=\fperc/5$ and
$\gamma=2.3$ are shown in Fig.~\ref{fig:stru_summ23} for different strategies.
Similar  results (data not shown) were found for $\gamma =2.8$ but with stronger
finite-size effects. In this regime, the largest connected component (LCC)
corresponds to approximately  99\%, 97\%, and 93\% for AcI-HP, AcI, and TgI,
respectively. The tail distribution $\Pi(k)$, defined as the probability that a
randomly chosen vertex has degree larger than $k$, decays very consistently with
a stretched exponential given by
\begin{equation}
\label{eq:fit_stre}
\Pi(k) = \sum_{k'\ge k} P(k') \sim \exp\left(-ak^{1/b}\right)
\end{equation}
with $b>1$ while $\Pi(k)\sim k^{-\gamma+1}$ is observed for the original
networks, as expected for SF degree distributions~\cite{Dorogovtsev2008}; see
Fig.~\ref{fig:stru_summ23}(b). The values of exponent $1/b$ obtained using
regressions to stretched exponential  for $k\ge 10$  for networks of size
$N=10^8$ immunized according to AcI were $1/b\approx 0.516$ and $0.42$ for
$\gamma=2.3$ and $2.8$, respectively. In the case of AcI-HP, the exponents
present smaller values $1/b\approx 0.185$ and $0.053$ for $\gamma=2.3$ and
$2.8$, respectively. All regressions provided correlation coefficients
$r\ge0.9995$. The degree of the most connected vertex is given by $N\Pi(\kmax)\sim
1$~\cite{Dorogovtsev2008} resulting $\kmax\sim (\ln N)^{b}$ for stretched
exponentials and $\kmax\sim N^{1/(\gamma-1)}$ for power-laws while a finite
upper rigid cutoff quickly appears in the case of TgI; see
Fig.~\ref{fig:stru_summ23}(c).

\begin{table}[hbt]
	\centering
	\setlength{\tabcolsep}{6pt}
	\begin{tabular}{cccc}
		\hline\hline
		& AcI-HP & AcI & TgI \\
		\hline
		$\gamma = 2.3$ & $0.50(1)$ & $0.46(1)$ & $0.20(1)$ \\
		$\gamma= 2.8$ & $0.55(1)$ & $0.51(1)$ & $0.24(1)$ \\
		\hline	\hline	
	\end{tabular}
	\caption{Percolation thresholds estimated from Fig.~\ref{fig:perc}. Numbers in
		parenthesis are the uncertainties in the last digit.}
	\label{tab:pth}
\end{table}

\begin{figure}[bht]
	\centering
	\includegraphics[width=0.85\columnwidth]{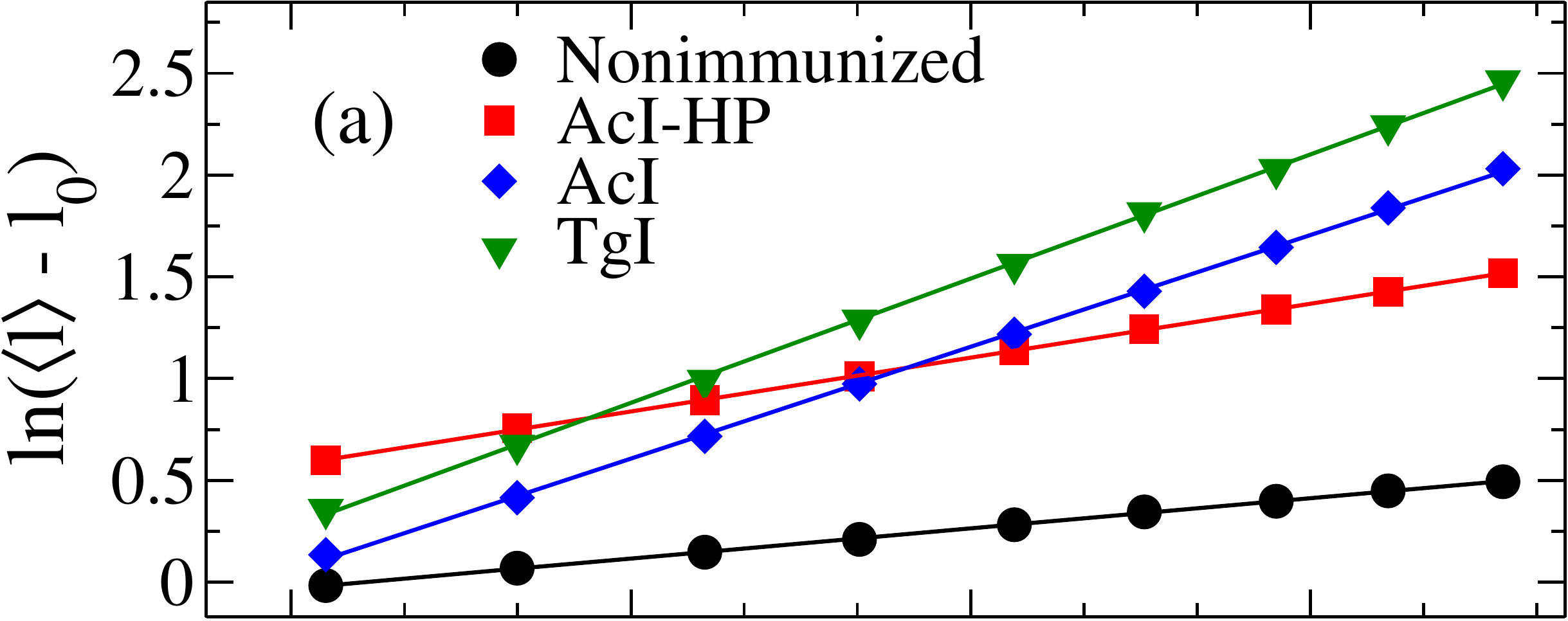}\\\vspace{0.041cm}
	\includegraphics[width=0.85\columnwidth]{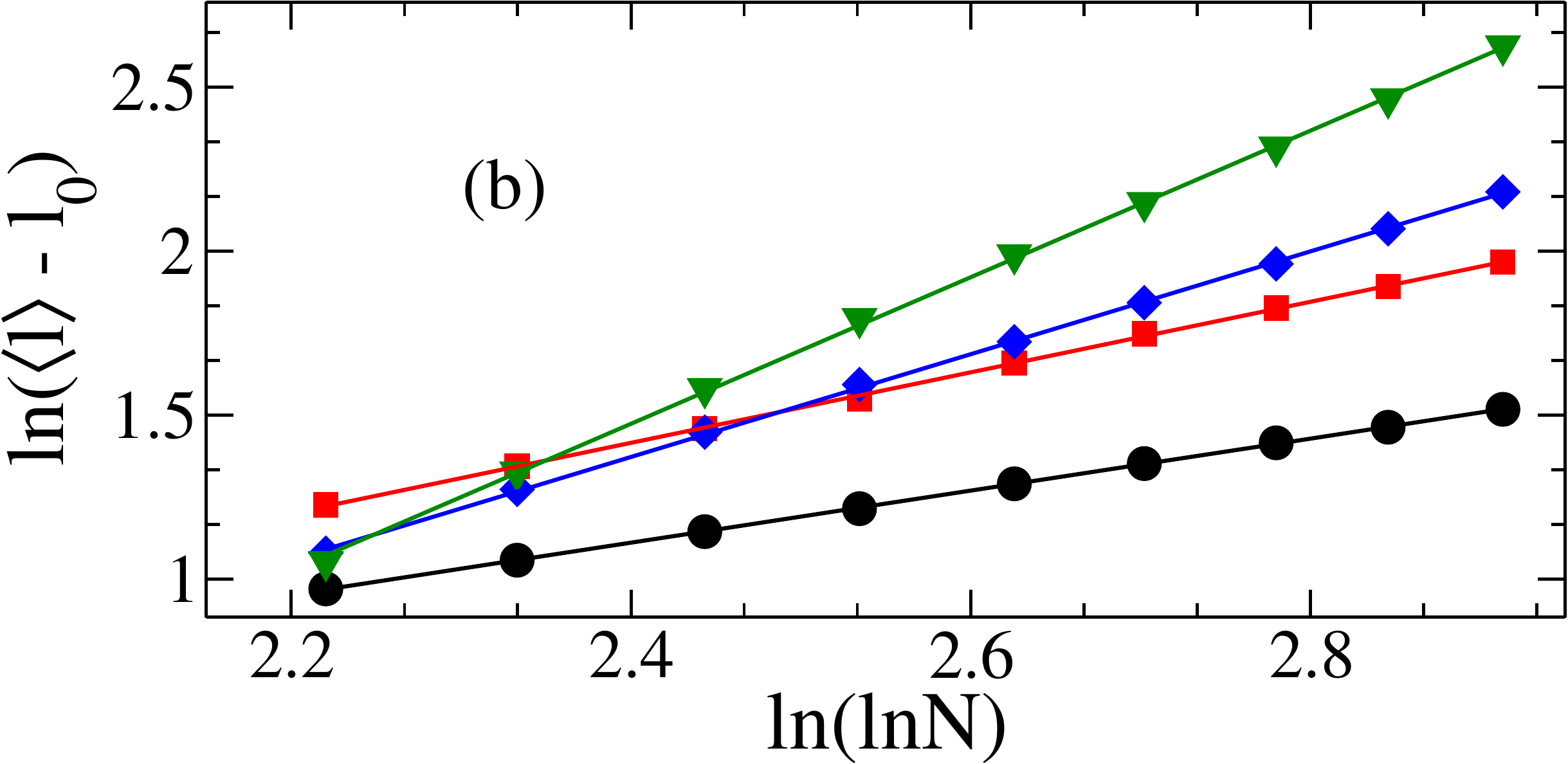}
	\caption{Finite-size scaling of the average shortest path using the ansatz of
		Eq.~\eqref{eq:el} for (a) $\gamma = 2.3$ and (b) $\gamma=2.8$. Symbols  
		represent simulations and solid lines regressions. Abscissas are the same 
		in both plots.}
	\label{fig:mpscal}
\end{figure}

Last but not least, Figure~\ref{fig:mpscal} shows the size dependence of the
average shortest paths calculated using  breadth first search
algorithm~\cite{Newman2014} for nonimmunized  and immunized UCM networks  with
$f=\fperc/5$. To verify if the immunized networks preserve the small-world
behavior, in which distances increase logarithmically with size, we fitted the
data to the expression
\begin{equation}
\av{l} = l_0 + C_0w^\alpha,
\label{eq:el}
\end{equation}
where $w=\ln N$. The ansatz given by Eq.~\eqref{eq:el} is not expected to work
exactly but to indicate a growth slower than  power-laws that is sufficient to
characterize the small-world property. All curves are very well fitted
(correlation coefficient $r\ge0.9998$) by Eq.~\eqref{eq:el} as shown in
Fig.~\ref{fig:mpscal}. The scaling exponents for $\gamma=2.3$ ($\gamma=2.8$ )
were $\alpha=1.32$, $2.77$, and $3.05$ ($\alpha=1.05$, $1.56$, and $2.33$) for
AcI-HP, AcI, and TgI, respectively. The values  larger than unity indicate a
super-logarithm growth for immunized networks and is larger for more efficient
immunization. For the nonimmunized networks, we found $\alpha<1$ which reflects
the sub-logarithm growth expected for random SF
networks~\cite{barabasi2016network}.

\section{Epidemic thresholds for immunized synthetic networks}
\label{sec:sinSIS}
We ran standard Markovian SIS dynamics~\cite{Pastor-Satorras2015} on the
immunized networks using the algorithm detailed in Ref.~\cite{Cota2017} and
summarized in Appendix \ref{app:compSIS}. We analyzed the steady-state regime
with quasistationary (QS) simulations~\cite{DeOliveira2005,Sander2016} to
circumvent the drawbacks of the absorbing states in finite
sizes~\cite{Marro2005}. We determined the epidemic threshold in stochastic
simulations using the position $\lambda_\text{c}$ of maximum of the dynamical
susceptibility $\chi(\lambda)$ defined as
$\chi=N[\av{\rho^2}-\av{\rho}^2]/\av{\rho}$~\cite{Ferreira2012}, $\rho$ being
the density of infected nodes and the averages are computed in the QS
regime~\cite{Cota2017,Sander2016}.  Appendix~\ref{app:QSan} gives some details
of the QS analysis.

\begin{figure}[tbh]
	\centering
	\includegraphics[width=0.9\linewidth]{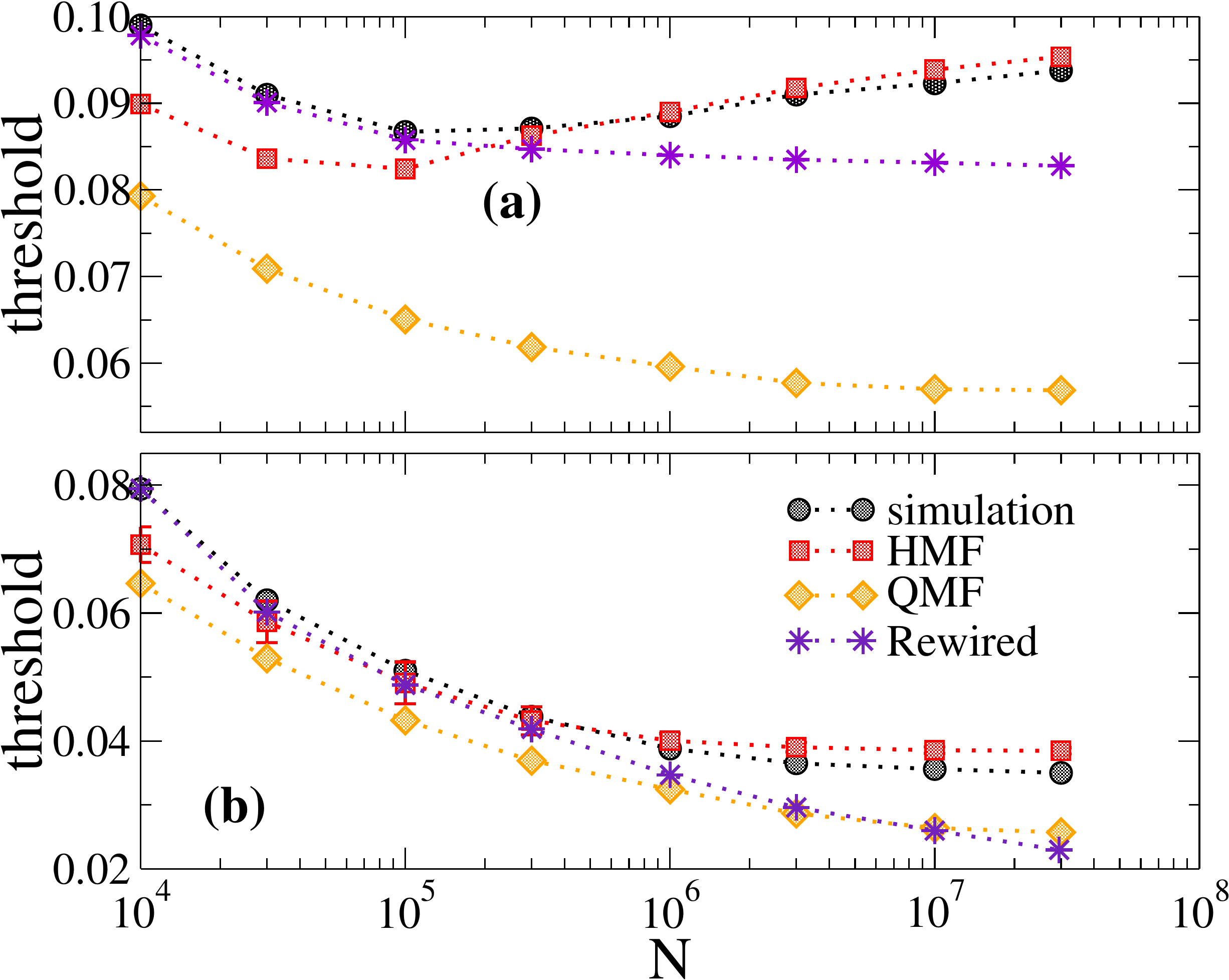}
	\caption{Simulation and mean-field epidemic thresholds  as functions of the
		system size for the SIS model on UCM networks with $\gamma=2.3$ and a fraction
		$f=\fperc/5\approx 0.1$ of vertices immunized using either (a) AcI or (b)
		AcI-HP methods. Stochastic simulations for immunized networks without (circles)
		and with (stars) degree-preserving rewiring are shown.Abscissas are the same in
		both plots.
		}
	\label{fig:lb_teor23AcI}
\end{figure}

\begin{figure}[hbt]
	\centering
	\includegraphics[width=0.9\linewidth]{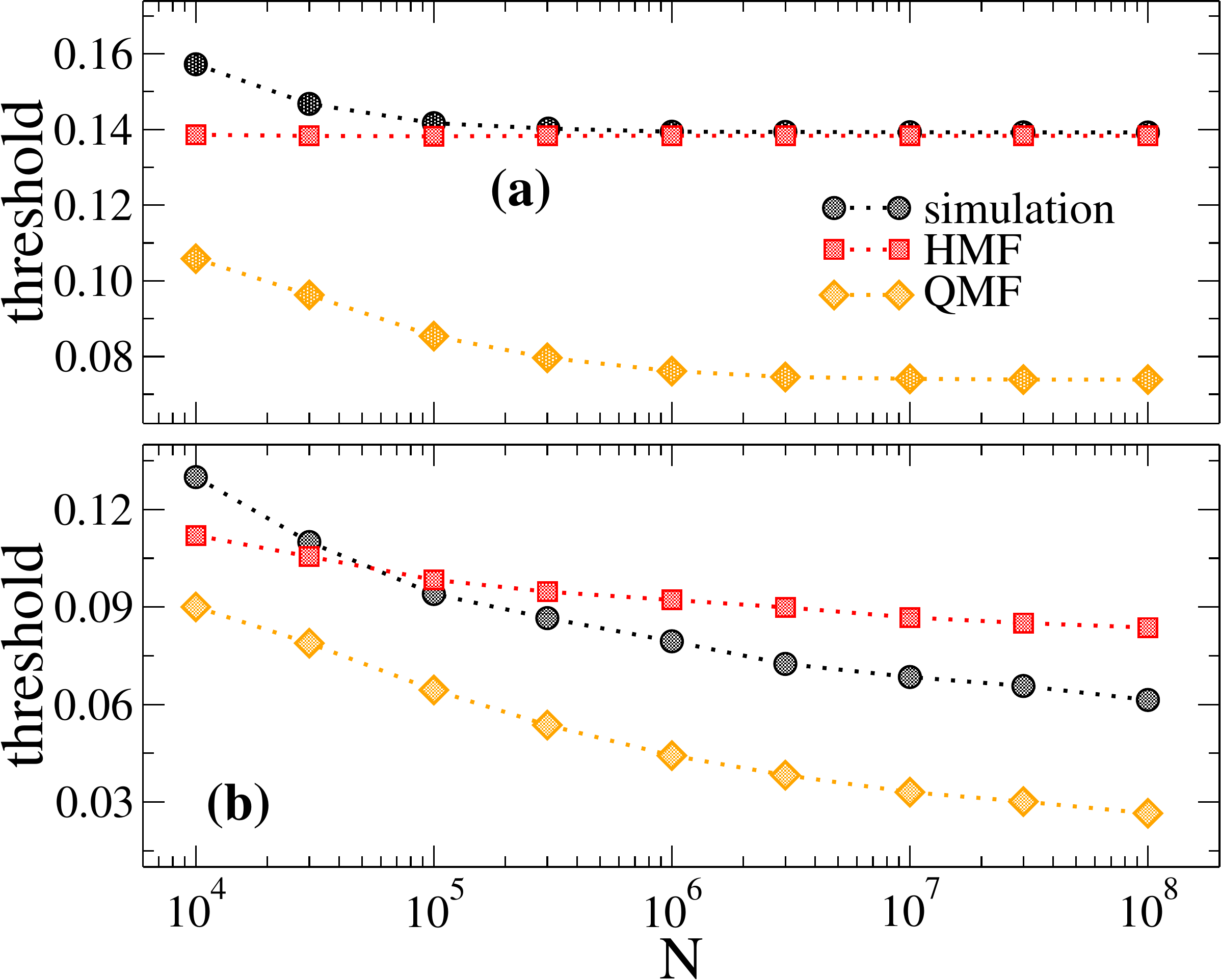}
	\caption{Simulation and mean-field epidemic thresholds  as functions of the
		system size for the SIS model on UCM networks with $\gamma=2.8$ with a fraction
		$f=\fperc/5\approx 0.1$ of vertices immunized using either (a) AcI  or (b)
		AcI-HP methods. Abscissas are the same in both plots. }
	\label{fig:fsslb28}
\end{figure}

For the synthetic SF networks, the epidemic threshold usually depends on
the effective size  of the LCC  of nonimmunized vertices. So, one cannot
investigate the finite-size effects independently of the immunization fraction
if we fix the size $N_0$ of the original network. Therefore, larger  networks
were generated such that the LCC lies in the range $[0.95N_0,1.05N_0]$.
Simulations are run on the LCC. The dependence of the  epidemic threshold with
size for UCM networks with $\gamma=2.3$ is shown in Fig.~\ref{fig:lb_teor23AcI}
for $f= \fperc/5 \approx 0.1$ using both AcI and AcI-HP methods. These
thresholds approach finite values as $N\rightarrow\infty$ in both cases
including the weakly supervised AcI-HP method. The AcI case shows the epidemic
thresholds increasing with size after an initial decay since the strategy
becomes more efficient as larger hubs appear in the networks. A similar behavior
cannot be discarded for AcI-HP for much larger sizes. Simulations for
$\gamma=2.8$ in Fig.~\ref{fig:fsslb28} are qualitatively similar but subject to
different finite-size effects. 

In Figs.~\ref{fig:lb_teor23AcI} and \ref{fig:fsslb28}, simulations are compared
with the  heterogeneous mean-field (HMF) theory~\cite{Pastor-Satorras2001}, which takes
in to account only the degree distribution, and the quenched men-field (QMF)
theory~\cite{Chakrabarti2008}, which includes the detailed network structure
through its adjacency matrix $A_{ij}$~\cite{Newman2014}. Details of the theories
are presented in Appendix \ref{app:mft}. The theoretical epidemic thresholds are
$\lambda_\text{c}^\text{HMF} = {\av{k}}/{\av{k^2}}$~\cite{Pastor-Satorras2001}
and $\lambda_\text{c}^\text{QMF}=1/\Lambda_1$~\cite{Chakrabarti2008}, where
$\Lambda_1$ is the largest eigenvalue of $A_{ij}$. Notice that HMF outperforms
the more detailed QMF theory in the case of immunized networks,  conversely to
the mean-field performances for the SIS model on nonimmunized networks~\cite{Silva2019}.

The asymptotically finite epidemic thresholds cannot be justified only by the pruning of
hub' degrees since AcI and, mainly, AcI-HP methods, lead to stretched tail
distributions  expected to asymptotically produce a null epidemic threshold according to
rigorous results~\cite{Huang2018}. We tackled this point performing a
degree-preserving rewiring of the effective immunized network and rerunning the
SIS process. Rewired networks for AcI-HP  have epidemic thresholds decaying with size, 
compatibly with QMF and consistent with the conjecture of a vanishing epidemic threshold
for a stretched exponential~\cite{Huang2018} whereas the nonrewired ones present
saturation consistent with the HMF theory. In the AcI case, the rewiring changes
the tendency of increasing to a very slow decay, still qualitatively compatible the
QMF theory but with a much larger prefactor.

\begin{figure}[hbt]
	\centering
	\includegraphics[width=0.9\columnwidth]{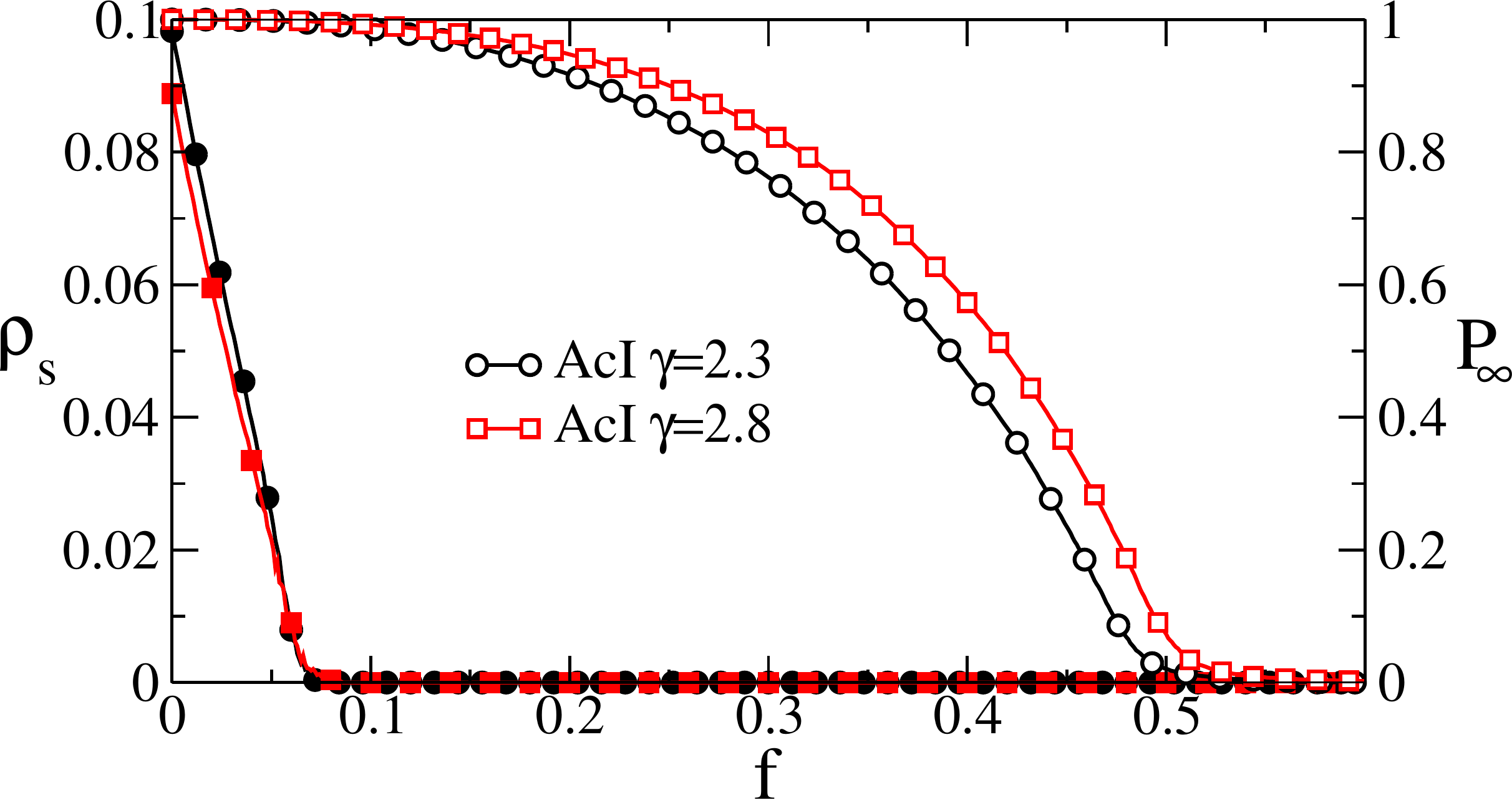}	
	\caption{Stationary density of infected vertices (closed symbols and left-hand
		axis) and relative size of the LCC (open symbols and right-hand axis) as
		functions of the fraction of vertices immunized	with the AcI method on UCM
		networks with $N=10^4$ and two degree exponents. The	infection rates are
		$\lambda_0=0.086$ and $0.145$  for $\gamma=2.3$	and 2.8, respectively.}
	\label{fig:rhoN10}
\end{figure}
A second fundamental ingredient for determining the SIS spreading in networks
with large degree exponents or stretched exponential tails is the typical
separation among central spreaders~\cite{Chatterjee2009, Mountford2013,
	Boguna2013, Ferreira2016a}. Average shortest distances increase after
immunization as shown in Fig.~\ref{fig:stru_summ23}(d). The mechanism for
sustaining SIS activity on this kind of network can be summarized as
follows~\cite{Boguna2013}: A hub stays active in isolation for long times
through a feedback mechanism where it infects its neighbors which in turn
reinfect the hub. If this time is long enough and distances among hubs increase
sufficiently slowly with size, rare fluctuations can promote the mutual
activation of hubs in the thermodynamical limit  even if they are not directly
connected, triggering an endemic  phase  and leading to an asymptotically null
epidemic threshold compatible with the QMF
theory~\cite{Huang2018,Boguna2013,Castellano2019}. Otherwise, a finite epidemic
threshold, compatible with HMF, would be
observed~\cite{Cota2018a,Ferreira2016a}. Indeed, acquaintance immunizations act
on both properties:  pruning  the hub' degrees, reducing their capabilities to
stay active, and increasing the distances among them, damping their mutual
interactions. The finite epidemic threshold  is consistent with a
collective~\cite{Cota2018a,Ferreira2016a} rather than specific-motif (hub,
maximum $k$-core, etc..) driven activation
mechanism~\cite{Kitsak2010,Castellano2012}. This outcome is remarkable since a
viable, adaptive, and weakly supervised strategy as the AcI-HP can efficiently
contain the onset of an endemic state for a very aggressive epidemic process (no
acquired immunity) as the SIS model.

\section{Percolation versus immunization thresholds on real networks}
\label{sec:threReal}

\begin{figure*}[hbt]
	\centering
	\begin{minipage}{0.452\linewidth}
		\includegraphics[width=0.999\linewidth]{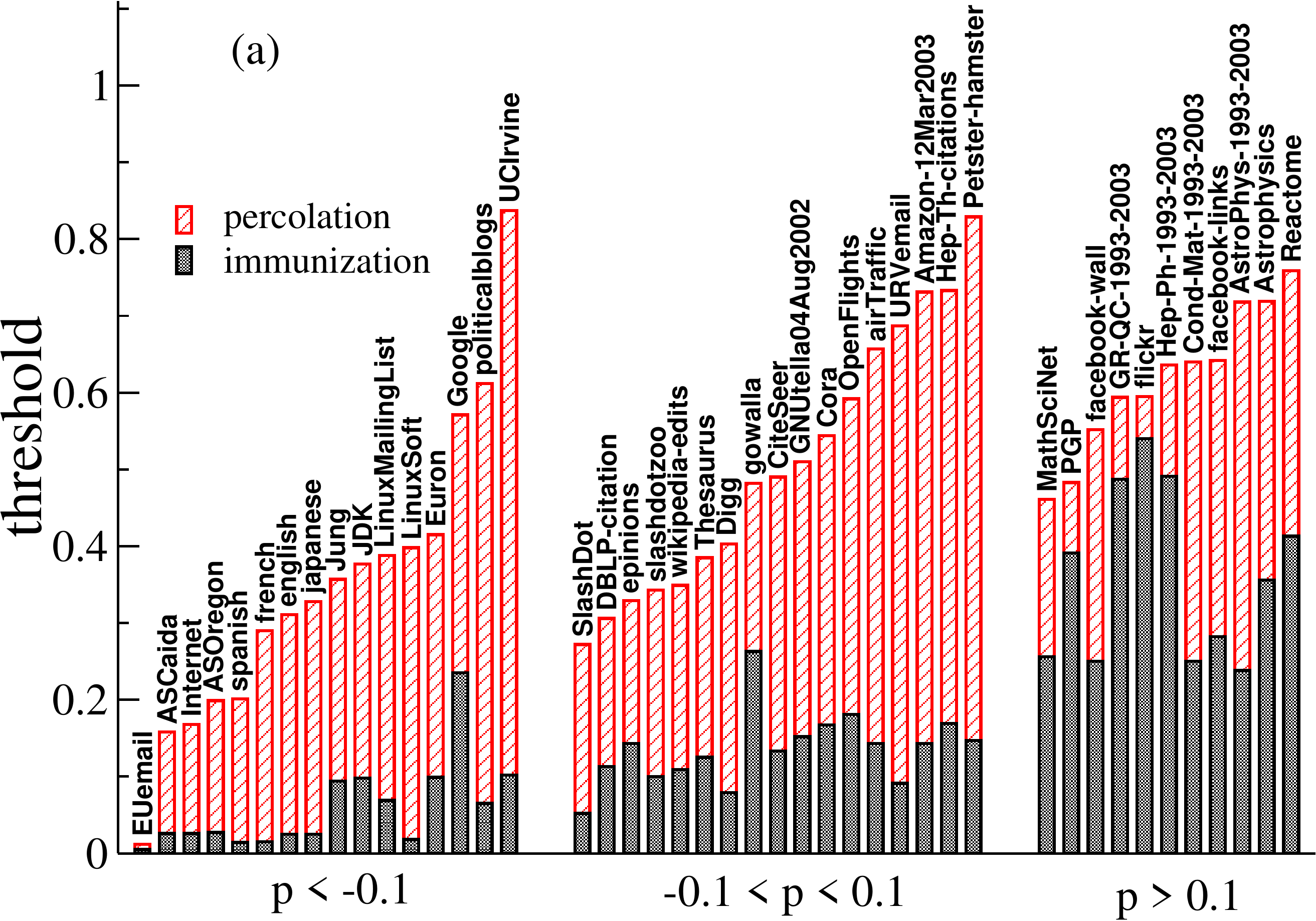}
	\end{minipage}~
	\begin{minipage}{0.451\linewidth}
		\includegraphics[width=0.999\linewidth]{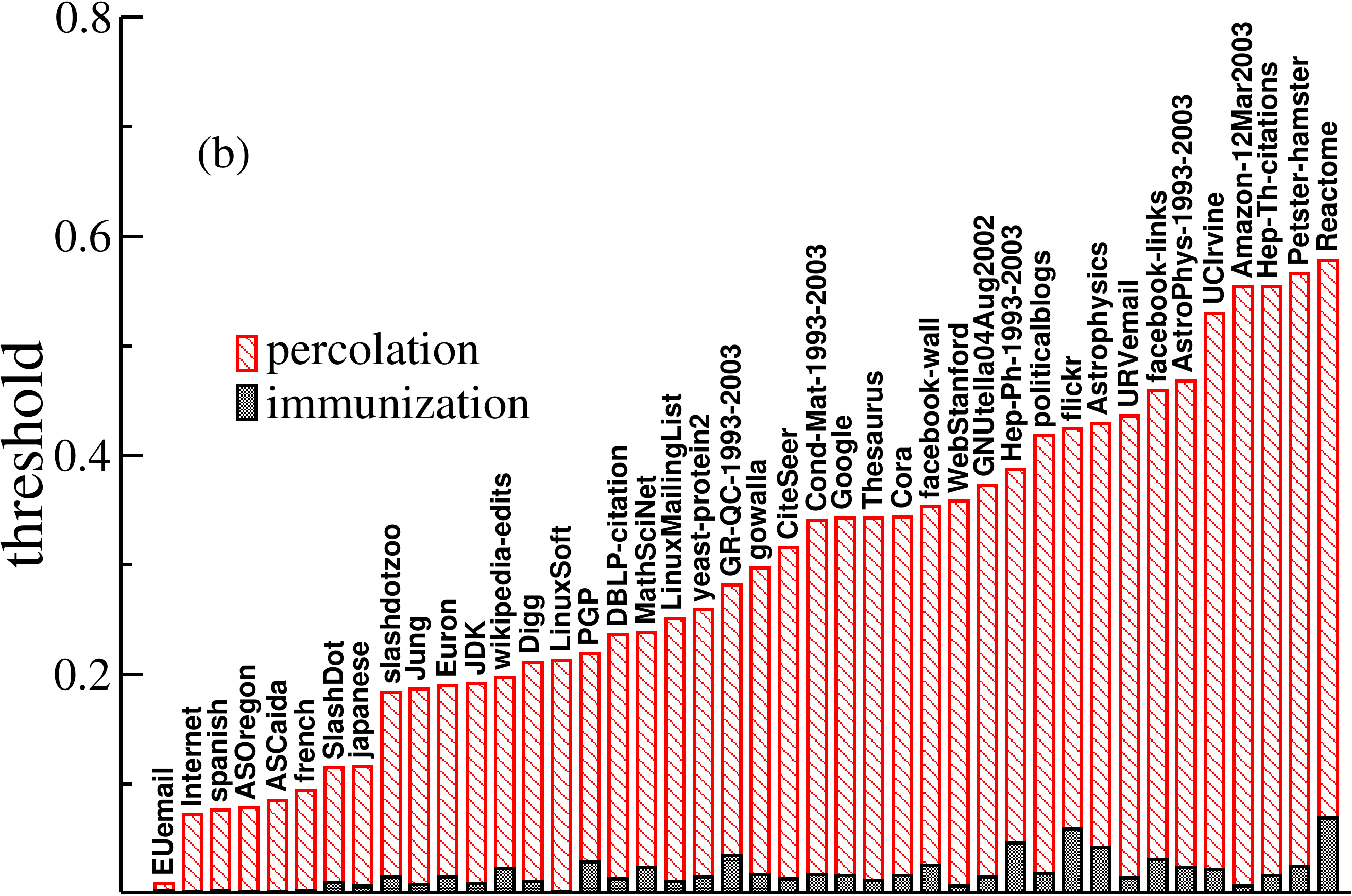}	
	\end{minipage}\\
    \begin{minipage}{0.409\linewidth}
    	\includegraphics[width=0.809\linewidth]{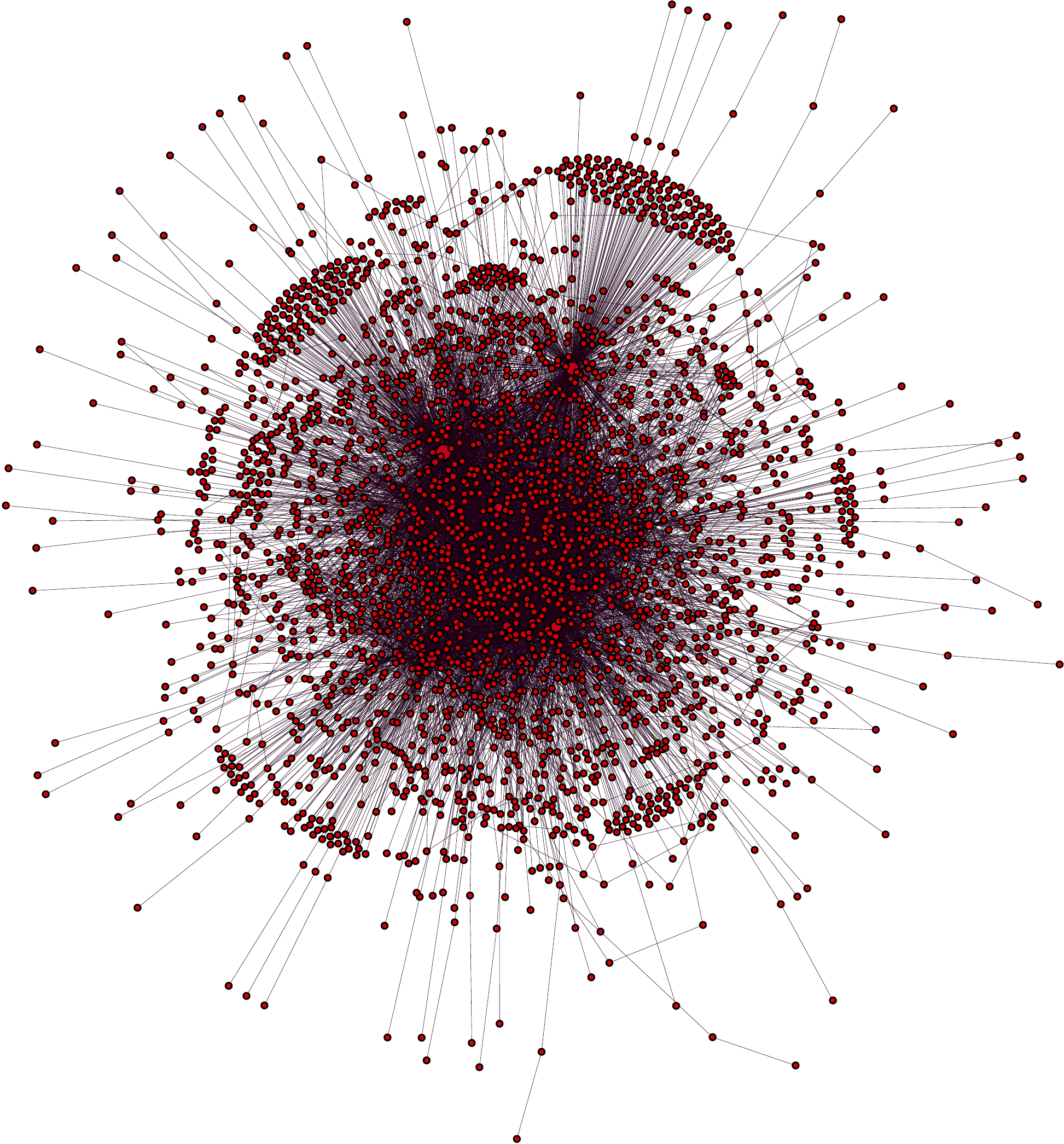}(c)
    \end{minipage}
    \begin{minipage}{0.409\linewidth}
       	\includegraphics[width=0.809\linewidth]{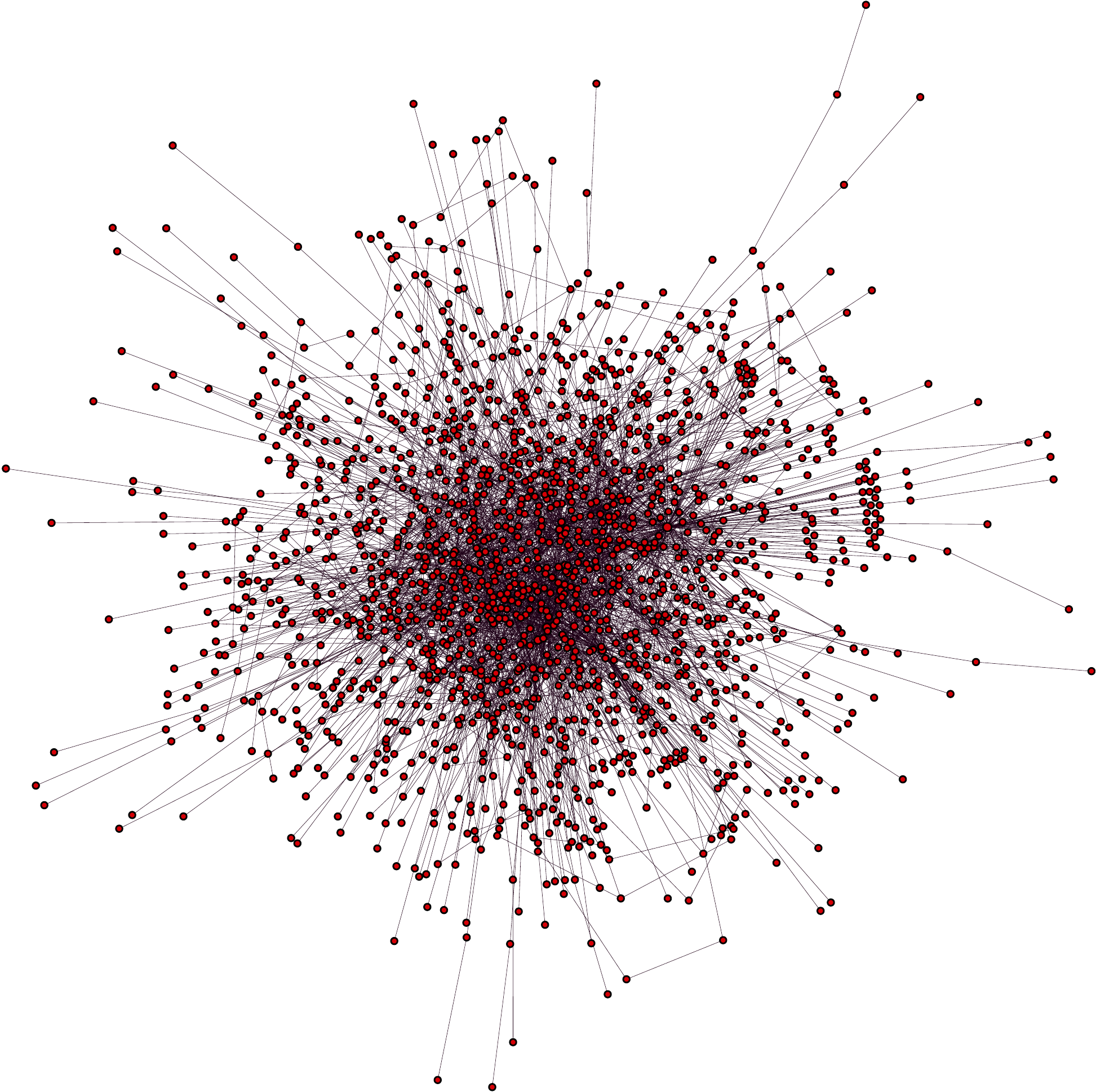}(d)
    \end{minipage}
\caption{Comparison of the immunization and percolation	thresholds for (a) AcI
	and (b)TgI  using $\rho_0=0.1$ for  a set of 42 real	networks. In (a) data are
	grouped according to the ranges of Pearson	coefficients  indicated in the
	abscissa. Averages were computed over $10^2$ independently realizations of
	immunization for each network in the case of AcI-HP while TgI is deterministic.
	Network representations of the LCC of the adjacency vocabulary networks for
	Japanese considering (c) nonimmunized  and (d) immunized  versions using AcI
	with $f=\fimmc$. 	Values of $\lambda_0$ used for real networks are given in
	Appendix
		\ref{app:strucreal}.}
	\label{fig:DynvsPercAcI}
\end{figure*}

The aforementioned impact of non massive immunization on synthetic networks
naturally calls for applications on real networks. So, we determined the
immunization fraction capable to eradicate a highly endemic steady state in  the
nonimmunized network~\cite{Pastor-Satorras2002,Matamalas2018}. We calculated the
infection rate for which the stationary density of infected vertices in the
nonimmunized networks is $\rho=\rho_0$ using standard simulations (without QS
sampling). Starting with $f=0$, we increase $f$ with small increments $\Delta
f\ll 1$ at infection rate  $\lambda=\lambda_0$, until the stationary density
drops to zero, determining the \textit{immunization threshold} $\fimmu$. For our
simulations, we used $\rho_0 = 0.1$. Results are qualitatively similar for other
values of $\rho_0$. Figure~\ref{fig:rhoN10} shows the stationary density and
relative size of the LCC as functions of $f$ for UCM networks using AcI. The
stationary density falls to zero far below  the percolation threshold,
$\fimmu\ll \fperc$, confirming the high efficacy of non massive immunization for
epidemic containment in synthetic networks.

We now turn our attention to  a set of 42 real networks of wide spectrum of
structural properties previously investigated in Ref.~\cite{Radicchi2015}; see
Appendix~\ref{app:strucreal} for basic network information. As in the
percolation analysis, we tackle the finite size of the networks assuming that
either a stationary density or a relative size of the LCC below $10^{-3}$
correspond to the immunization or percolation thresholds, respectively. Again,
the results depend little on this choice. Visualizations of the  LCC of the
adjacency vocabulary network for Japanese and its immunized version using AcI at
$f=f^\text{imm}_\text{c}=0.024$ are shown in Fig.~\ref{fig:DynvsPercAcI}. The
immunized network presents a sparser but well connected LCC having more than
70\% of the vertices and concentrated in the innermost regions of the network.
Considering only the LCC, the average degree decays from $\av{k}=5.92$ to $3.47$
while the average distance increases from $\av{\ell}=3.07$ to $4.26$ when AcI is
applied. The percolation and immunization thresholds are $\fperc=0.32$ and
$\fimmu=0.024$, respectively.

Figure~\ref{fig:DynvsPercAcI}(a) compares the immunization and
percolation thresholds  calculated using AcI for the set of 42 real networks. The
data is grouped according to the Pearson coefficient $p$~\cite{Newman2002} 
defined as~\cite{Newman2014}
\begin{equation}
p = \frac{\sum_{ij}\left(A_{ij}-\frac{k_ik_j}{N\av{k}}\right)k_ik_j}
{\sum_{ij}\left(k_i\delta_{ij}-\frac{k_ik_j}{N\av{k}}\right)k_ik_j},
\label{eq:pearson}
\end{equation}
which lays in the interval $-1<p<1$ and ranks the level of degree correlations
of the network, being disassortative for $p<0$, uncorrelated for $p=0$ and
assortative for $p>0$; see Appendix~\ref{app:strucreal} for the Pearson coefficients
and  values of $\lambda_0$ for $\rho_0=0.1$ of the real networks. The efficiency
of AcI  is negatively correlated with the Pearson coefficient with a
immunization threshold much lower than the percolation one for both highly
disassortative ($p<-0.1$) and slightly correlated ($|p|<0.1$) cases and a worse
performance for the highly assortative cases ($p>0.1$). Such a dependence
reflects the loss of efficiency of acquittance-based methods for finding hubs on
assortative networks. Our results are in agreement with
Ref.~\cite{Gomez-Gardenes2006} where it was observed that immunization
efficiency depends on the level of correlation.

Much better performances are attainable if cleverer immunization strategies are
adopted. We investigated  TgI in Fig.~\ref{fig:DynvsPercAcI}(b) which is much
more efficient than AcI. The condition $\fimmu\ll \fperc$ holds for  the whole
set of real networks. One can  improve further  with more specific centralities
rather than degree~\cite{Kitsak2010} or use process-targeted
strategies~\cite{Matamalas2018}. However, whatever the used approach one must
rely on the dynamic processes  rather than only topological structures.

\begin{figure}[hbt]
	\centering
	\includegraphics[width=0.9865\linewidth]{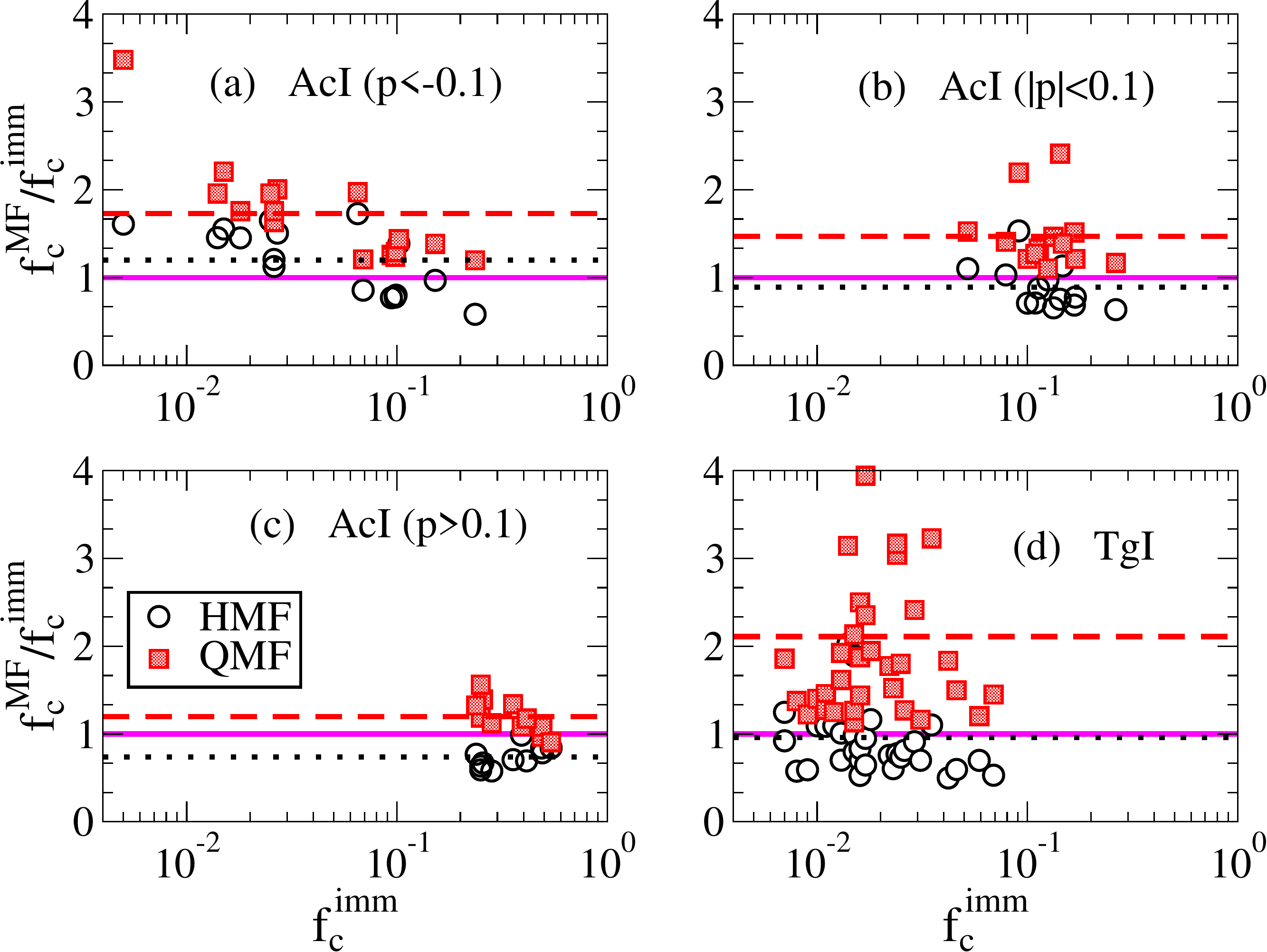}
		\caption{Comparison of the ratio between  immunization thresholds obtained in
			numerical simulations and  mean-field theories  for a set of 42 real networks
			immunized using either (a)-(c) AcI and (d) TgI strategies with $\rho_0=0.1$.
			Data for AcI are grouped according ranges of Pearson coefficient: (a)
			$p<-0.1$, (b) $|p|<0.1$, and (c) $p>0.1$. Dashed and dotted lines represent
			the ratios averaged over all networks for QMF and HMF theories, respectively.
			Averages as in Fig.~\ref{fig:DynvsPercAcI}.}
\label{fig:dynamical_teor}
\end{figure}

We conclude our results comparing the simulation values of immunization thresholds 
with two mean-field theories. Within a HMF theory, $\fimmc$ is given by the condition
\begin{equation}
\frac{\av{k^2}_{\fimmc}}{\av{k}_{\fimmc}}=\lambda_0,
\end{equation} 
where the $\av{k^n}_f$ are moments of the
degree distribution of the LCC after immunization of $fN$
vertices.
Similarly, the QMF immunization threshold is given by 
\begin{equation}
\frac{1}{\Lambda_1({\fimmc})}=\lambda_0,
\end{equation} where $\Lambda_1(f)$ is the largest eigenvalue of the adjacency
matrix corresponding to the LCC after immunization of a fraction $f$. The ratio
$f_\text{c}^\text{MF}/\fimmc$ between theory and simulation for immunization
thresholds  on real networks are compared in Fig.~\ref{fig:dynamical_teor} for
AcI and TgI strategies. The close the ratio is to 1 (solid lines) the better is
the mean-field theory.  The HMF theory tends to underestimate while QMF to
overestimate the immunization thresholds in opposition to the performance of the
epidemic thresholds where QMF tends to underestimate and HMF to overestimate the
simulation results~\cite{Silva2019}. For efficient immunizations, namely
TgI and AcI at disassortative or weakly correlated networks, the HMF theory
performs better than QMF. The cases with highly assortative correlations, the
performances are similar. This can be seem with the aid of  average value  of
the ratios computed over each plot and indicated by dashed (QMF) and dotted
(HMF) horizontal lines, respectively.

\section{Conclusions}
\label{sec:conclu}
Containment methods for controlling  propagation of dynamical processes on the
top of networks is crucial for setting up protection protocols against
threatenings that can be disseminated throughout networked substrates. A
considerable part of the containment methods are  based on percolation analysis
while the spreading on partially and weakly damaged networks has received little
attention. In the present work, we tackle this problem investigating the
epidemic spreading of the SIS model on complex networks using different
immunization strategies. We  report that a non massive immunization with the
removal of a fraction far below the percolation threshold can alter the
originally motif-driven (hubs, maximum $k$-core, etc...) \cite{Kitsak2010}
mechanisms for activation of endemic phases to a collective activation involving
extensive parts of the network. Even in the case of a weakly supervised
immunization strategy, the absence of an epidemic threshold at originally SF
networks is replaced by finite thresholds caused by the concomitant pruning of
hubs and increasing of their mutual distances. Backed up by the analysis of a
collection of real networks, we also show that immunization can efficiently
contain epidemic spreading using non massive levels.

To the best of our knowledge, the information that immunization thresholds are
much smaller than the percolation ones has passed unnoticed or underestimated in
the vast physics literature concerned with immunization of complex networks. So,
we hope that our work will ignite new research activity towards elaboration of
optimal and viable immunization strategies. We conclude highlighting the
importance of running accurate stochastic simulations of the actual dynamical
processes since the long-range interactions cannot be completely reckoned by
mean-field methods~\cite{Silva2019,Huang2018}.


\begin{acknowledgments}
	This work was partially supported by the Brazilian agencies CNPq and FAPEMIG.
	This study was financed in part by the Coordenação de Aperfeiçoamento de
	Pessoal de Nível Superior - Brasil (CAPES) - Finance Code 001.
\end{acknowledgments}

\appendix

\section{Structural properties of real networks}
\label{app:strucreal}
Tables~\ref{tab:propreal1}, \ref{tab:propreal2}, and \ref{tab:propreal3} show
some structural properties of the real networks studied on this work such as the
number of vertices $N$, mean degree $\av{k}$, heterogeneity coefficient $\eta =
\av{k^2}/ \av{k}$, and Pearson coefficient $p$. Different tables correspond to
different rages of Pearson coefficient used in Fig.~\ref{fig:DynvsPercAcI}(a).
The infection rate $\lambda_0$ necessary to sustain a stationary state with
$10\%$ of infected vertices in the original network are shown in the last
columns.

\begin{table}[hbt]
	\centering
	\setlength{\tabcolsep}{4pt}
	\begin{tabular}{lccccc}
		\hline\hline
		Network              &   N     & $\av{k}$ & $\eta $  & $p$      &$\lambda_0$\\
		\hline
		spanish              &$11,558$ &$7.45$   &$457      $&$-0.28$  &$0.0424   $\\
		japanese		     & $2,698$ &$5.92$   &$108      $&$-0.26$  &$0.0624   $\\
		english	             &$7,377$  &$12.0$   &$320      $&$-0.24$  &$0.0287   $\\
		french	             &$8,308$  &$5.73$   &$218      $&$-0.23$  &$0.0599   $\\
		Jung 			     &$6,120  $&$16.4    $&$991     $&$-0.23$  &$0.0225   $\\
		JDK 			     &$6,434  $&$16.7    $&$982     $&$-0.22$  &$0.0224   $\\
		politicalblogs	     &$1,222  $&$27.5    $&$81.2    $&$-0.22$  &$0.0203   $\\
		Internet     	     &$22,963 $&$4.21    $&$261     $&$-0.20  $&$0.0899   $\\
		ASCaida      	     &$26,475 $&$4.03    $&$280     $&$-0.19  $&$0.0924   $\\
		EUmail      	     &$224,832$&$3.02    $&$567     $&$-0.19  $&$0.0949   $\\
		UCIrvine      	     &$1,893  $&$14.6    $&$55.6    $&$-0.19  $&$0.0349   $\\
		LinuxMailingList     &$24,567 $&$12.9    $&$341     $&$-0.19  $&$0.0349   $\\
		ASOregon             &$6,474  $&$3.89    $&$165     $&$-0.18  $&$0.103   $\\
		LinuxSoft            &$30,817 $&$13.8    $&$853     $&$-0.18  $&$0.0274   $\\
		Google               &$15,763 $&$18.9    $&$902     $&$-0.12  $&$0.0224   $\\ 
		Euron                &$33,696 $&$10.7    $&$142     $&$-0.12  $&$0.0412   $\\
		\hline	\hline	
	\end{tabular}
	\caption{Some properties of real networks with Pearson coefficients $p<-0.1$.
		Size $N$, average degree $\av{k}$, heterogeneity coefficient $\eta$, and
		infection rate $\lambda_0$ able to produce a steady state with 10\% of infected
		vertices are shown.}
	\label{tab:propreal1}
\end{table}

\begin{table}[hbt]
	\centering
	\setlength{\tabcolsep}{4pt}
	\begin{tabular}{lccccc}
		\hline\hline
		Network              &   N     & $\av{k}$ & $\eta $  & $p$   &$\lambda_0$\\
		\hline
		PetsterHamster       &$1,788  $&$13.9   $&$45.5    $&$-0.089  $&$0.0393   $\\
		SlashDotZoo          &$79,166 $&$11.8   $&$146     $&$-0.075  $&$0.0349   $\\
		Wikipedia-edits      &$113,123$&$35.8   $&$689     $&$-0.065  $&$0.0112   $\\
		CiteSeer             &$365,154$&$9.43   $&$48.4    $&$-0.063  $&$0.0599   $\\
		Cora                 &$23,166 $&$7.69   $&$23.6    $&$-0.055  $&$0.0849  $\\	 
		Thesaurus            &$23,132 $&$25.7   $&$103     $&$-0.048  $&$0.0187   $\\  
		DBLP-citations       &$12,496 $&$7.93   $&$43.7    $&$-0.046  $&$0.0637   $\\
		Epinions             &$75,877 $&$10.7   $&$183     $&$-0.041  $&$0.0437   $\\
		SlashDot             &$51,083 $&$4.56   $&$81.5    $&$-0.035  $&$0.0899   $\\
		Hep-Th-citations     &$27,400 $&$25.7   $&$106     $&$-0.030  $&$0.0218   $\\
		gowalla              &$196,591$&$9.67   $&$306     $&$-0.029  $&$0.0524   $\\    
		Amazon12Mar2003      &$400,727$&$11.7   $&$30.3    $&$-0.020  $&$0.0699   $\\
		Gnutella04Aug2002    &$10,876 $&$7.35   $&$13.9    $&$-0.013  $&$0.101   $\\  
		Digg                 &$29,652 $&$5.72   $&$28.0    $&$0.003   $&$0.0874   $\\
		OpenFlights          &$2,905  $&$10.8   $&$55.8    $&$0.049   $&$0.0424   $\\
		URVemail             &$1,133  $&$9.62   $&$18.6    $&$0.078   $&$0.0734   $\\
		\hline	\hline	
		
	\end{tabular}
	\caption{Some properties of real networks with Pearson coefficients $|p|<0.1$.
		Quantities as defined in Table~\ref{tab:propreal1}.}
	\label{tab:propreal2}
\end{table}

\begin{table}[hbt]
	\centering
	\setlength{\tabcolsep}{4pt}
	\begin{tabular}{lccccc}
		\hline\hline
		Network              &   N     & $\av{k}$ & $\eta $  & $p$   &$\lambda_0$\\
		\hline
		MathSciNet           &$332,689$&$4.93    $&$16.4      $&$0.10   $&$0.127   $\\
		Cond-Mat1993-2003    &$21,363 $&$8.55    $&$22.4      $&$0.13   $&$0.0787   $\\
		facebook-links       &$63,392 $&$25.8    $&$88.0      $&$0.18   $&$0.0212   $\\
		AstroPhys1993-2003   &$17,903 $&$22.0    $&$65.7      $&$0.20   $&$0.0256   $\\
		facebook-wall        &$43,953 $&$8.29    $&$24.7      $&$0.22   $&$0.0724   $\\
		Astrophysics         &$14,845 $&$16.1    $&$45.5      $&$0.23   $&$0.0374   $\\
		PGP                  &$10,680 $&$4.55    $&$18.9      $&$0.24   $&$0.155   $\\
		Reactome             &$ 5,973 $&$48.8    $&$143       $&$0.24   $&$0.0117   $\\
		flickr               &$105,722$&$43.8    $&$349       $&$0.25   $&$0.0162   $\\
		Hep-Ph-1993-2003     &$11,204 $&$21.0    $&$131       $&$0.63   $&$0.0324   $\\
		GR-QC-1993-2003      &$4,158  $&$6.45    $&$18.0      $&$0.64   $&$0.130   $\\
		\hline	\hline	
	\end{tabular}
	\caption{Some properties of real networks with Pearson coefficients $p>0.1$.
		Quantities as defined in Table~\ref{tab:propreal1}.}
	\label{tab:propreal3}
\end{table}

\section{Numerical methods}
\label{app:numer}
\subsection{Computer implementation of SIS}
\label{app:compSIS}
To simulate the SIS model on graphs, we used the optimized Gillespie algorithm
(OGA), described in Ref.~\cite{Cota2017}.  We determine  the number of infected
vertices $N_\text{i}$ and their total number of edges $N_\text{e}$. At each time
step, one of the events healing or infection attempt is chosen with
probabilities $p = N_\text{i}/(N_\text{i} + \lambda N_\text{e})$ and $1-p$,
respectively. In the former, one infected vertex is selected at random and 
become susceptible.  In the latter, one infected vertex is selected  with
probability proportional to its degree and one of its nearest-neighbors is
chosen with equal chance. If the selected neighbor is susceptible, it becomes
infected. Otherwise, the simulation proceeds to the next step without change of
configuration. At the end of this infection/healing process, the time is
incremented by $ \Delta t = 1/(N_\text{i} + \lambda N_\text{e})$ while
$N_\text{i}$ and $N_e$ are updated accordingly. This  process is iterated until
the predetermined simulation time is reached.

\subsection{Quasistationary analysis}
\label{app:QSan}
The QS analysis is a method to investigate dynamical processes with
absorbing states as the SIS model. It consists of evaluating averages only over
samples that did not visit the absorbing states~\cite{Marro2005}. For
subcritical and critical simulations the dynamics falls very often into the
absorbing state resulting in short and noisy intervals of stationary data.
Dickman and de Oliveira \cite{DeOliveira2005} proposed a method to overcome this
problem where the dynamics jumps to an active configuration previously visited
along the evolution of the process every time the system falls into the absorbing
state.  Computationally,  configurations visited  during the simulation are stored
and constantly updated. One of them is randomly selected to restart the
simulation every time the absorbing state is visited.
\begin{figure}[tbh]
	\centering
	\includegraphics[width=0.8\linewidth]{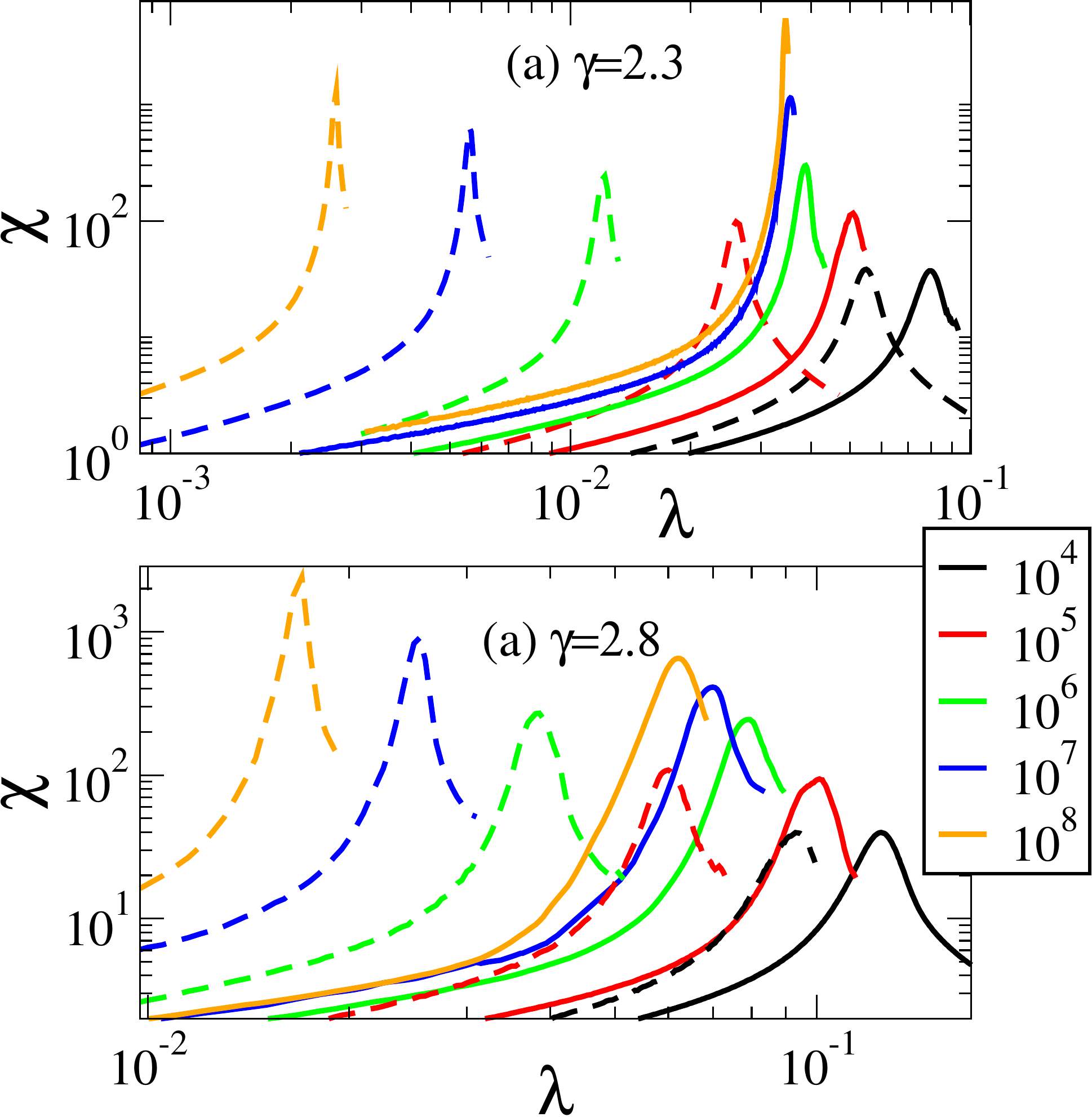}
	\caption{Susceptibility curves of the SIS model for immunized (solid lines)
		and nonimmunized (dashed lines) UCM networks with (a) $\gamma=2.3$ and (b)
		$\gamma=2.8$ and different sizes indicated in the legends. The immunization
		strategy is AcI-HP with $f=0.1$ in both cases.}
	\label{fig:sus}
\end{figure}

In our simulations, we started with all nodes infected and relax the system for
a time interval $t_\text{rlx} = 10^7$. The QS probability $Q(n)$
that the system has $n$  active (infected) vertices are computed  during the
time interval $t_\text{av} = 3\times 10^7$. A list with $M=100$ active
configurations is built. With probability $p_\text{r} = 0.01$ per time unit, the
list is updated replacing one of the configurations by the current state.
The QS density  and dynamical susceptibility are defined in terms of
moments
\begin{equation}
\av{\rho^s} = \frac{1}{N^s}\sum_{n\ge 1}n^sQ(n) 
\end{equation}
as $\rho_\text{qs} = \av{\rho}$ and $\chi = N
(\av{\rho^2}-\av{\rho}^2)/\av{\rho}$, respectively. Figure~\ref{fig:sus} shows
typical susceptibility curves for immunized and nonimmunized UCM networks. The
epidemic threshold is estimated as the position of maximum  susceptibility
value.

\section{Mean-field theories for the SIS model}
\label{app:mft}
\subsection{HMF theory}

The probability $\rho_k$ that a vertex of degree $k$ is infected in the HMF
theory evolves as
\begin{equation}
\dfrac{d \rho_k}{dt} = -\rho_k + \lambda k(1- \rho_k) \sum_{k'}P(k'|k)\rho_{k'},
\end{equation}
where $P(k'|k)$ is the conditional probability that a vertex of degree $k$ is
connected to a vertex of degree $k'$. Using linear stability analysis around the absorbing state 
$\rho_k=0$, the
following Jacobian is found $J_{kk'} = -\delta_{kk'}+\lambda k  P(k'|k)$, which
provide the epidemic threshold when its largest eigenvalue is zero. Considering 
uncorrelated networks where $P(k'|k) = \frac{k P(k')}{\av{k}}$  the epidemic
threshold of the SIS model becomes $\lambda_\text{c}^\text{HMF} =
\frac{\av{k}}{\av{k^2}}$~\cite{Pastor-Satorras2001}. Taking the asymptotic limit
of the moments $\av{k^n}$ we obtain
\begin{equation}
\lambda_\text{c}^\text{HMF} = \frac{\av{k}}{\av{k^2}}\simeq 
\begin{cases}
\kmax^{-3+\gamma} & \text{ if } \gamma<3 \\
\text{const.}    & \text{ if } \gamma>3
\end{cases},
\end{equation}
which goes to zero for $\gamma<3$ and becomes larger than zero for $\gamma>3$.

\subsection{QMF theory}
The QMF theory includes the network structure  by explicitly using the adjacency
matrix $A_{ij}$. The probability that a given vertex $i$  is infected evolves as
\begin{equation}
\dfrac{d \rho_i}{dt} = -\rho_i + \lambda (1- \rho_i) \sum_{j}A_{ij} \rho_j.
\end{equation}
The linear stability analysis around $\rho_i=0$ leads  to a Jacobian
matrix~\cite{Chakrabarti2008} $J_{ij} = -\delta_{ij} + \lambda  A_{ij}$, such
that the epidemic threshold is given by $\lambda_\text{c}^\text{QMF} =
{1}/{\Lambda_1}$, where $\Lambda_1$ is the largest eigenvalue of $A_{ij}$.
Plugging the expression for $\Lambda_1$ of uncorrelated networks derived in
Ref.~\cite{Chung2003}, one  obtains the following behavior for the  epidemic
threshold~\cite{Castellano2010}
\begin{equation}
\lambda_\text{c}^\text{QMF}  = \dfrac{1}{\Lambda_1} \simeq 
\begin{cases}
\av{k}/\av{k^2} & \text{if } 2 < \gamma < 5/2 \\
1/\sqrt{\kmax} & \text{if } 5/2 < \gamma 
\end{cases},
\end{equation}
which goes to zero for any power-law degree distribution irrespective of the
value of $\gamma$.


%

\end{document}